\let\accentvec\vec
\documentclass{llncs}

\let\vec\accentvec
\usepackage{amsmath}
\usepackage{amssymb}
\usepackage[dvips]{graphicx}
\usepackage{setspace}
\usepackage{epsfig}
\usepackage{subfigure}
\usepackage{caption,keywords}
\usepackage{color}
\usepackage{lipsum}
\usepackage{algorithm,algpseudocode}
\usepackage[T1]{fontenc}

\begin{document}

\title{Effective Capacity in Broadcast Channels with Arbitrary Inputs}
\author{Marwan Hammouda, Sami Akin, and J\"{u}rgen Peissig}
\institute{Institute of Communications Technology, Leibniz Universit\"{a}t Hannover
\email{\{marwan.hammouda, sami.akin, and peissig\}@ikt.uni-hannover.de}
\thanks{This work was partially supported by the European Research Council under Starting Grant--306644.}}

\maketitle

\begin{abstract}
We consider a broadcast scenario where one transmitter communicates with two receivers under quality-of-service constraints. The transmitter initially employs superposition coding strategies with arbitrarily distributed signals and sends data to both receivers. Regarding the channel state conditions, the receivers perform successive interference cancellation to decode their own data. We express the effective capacity region that provides the maximum allowable sustainable data arrival rate region at the transmitter buffer or buffers. Given an average transmission power limit, we provide a two-step approach to obtain the optimal power allocation policies that maximize the effective capacity region. Then, we characterize the optimal decoding regions at the receivers in the space spanned by the channel fading power values. We finally substantiate our results with numerical presentations.
\end{abstract}

\section{Introduction}
Cooperative communications can provide promising solutions to satisfy the ever-increasing demand for wireless data transmission \cite{tao2012overview}. Therefore, it has been investigated from several perspectives. For instance, the authors have explored the communication throughput in broadcast channels by invoking information-theoretic tools in \cite{cover1972broadcast,bergmans1973random,gallager1974capacity,jindal2004duality,tse1997optimal,li2001capacity,gupta2006power}. Particularly, considering one transmitter and two receivers, Cover obtained the achievable rate regions \cite{cover1972broadcast}. Then, this scheme was generalized to broadcast channels with many receivers in \cite{bergmans1973random}. Furthermore, the authors in \cite{li2001capacity} defined the ergodic capacity regions for fading broadcast channels considering different spectrum-sharing techniques and derived the optimal resource allocation policies that maximize these regions. Besides, the authors examined parallel Gaussian broadcast channels and obtained the optimal power allocation policies that achieve any point on the capacity region boundary subject to a sum-power constraint \cite{gupta2006power}.

In the aforementioned studies, the authors considered Gaussian input signaling. On the other hand, it is known that many practical systems make use of input signaling with discrete and finite constellation diagrams. In that regard, the authors in \cite{gunduz2010gaussian} studied two-user broadcast channels with arbitrary input distributions subject to an average power constraint and derived the optimal power allocation policies that maximize the weighted sum rate in low and high signal-to-noise ratio regimes. Similarly, the authors in \cite{lozano2006optimum} considered the mutual information in parallel Gaussian channels and derived the optimal power allocation policies. In addition, the authors in \cite{nguyen2010outage} explored the optimal power policies that minimize the outage probability in block-fading channels when arbitrary input distributions are applied under both peak and average power constraints. In these studies, the authors benefited from the fundamental relation between the mutual information and the minimum mean-square error (MMSE), which was initially established in \cite{guo2005mutual}. 

In another line of research, cross-layer design concerns gained an increasing interest since many of the current wireless systems are to support delay-sensitive applications. Consequently, quality-of-service (QoS) requirements in the form of delay and buffer overflow were studied in wireless communications from data-link and physical layer perspectives. Effective capacity was proposed as a performance metric that provides the maximum constant data arrival rate at a transmitter buffer that can be supported by a given service (channel) process \cite{wu_negi}. Subsequently, effective capacity was scrutinized in several different communication scenarios \cite{tang,musavian2010effective,gursoy,elalemmoh,ak_gur_4,ozcan2014qos}. For instance, effective capacity was examined in one-to-one transmission scenarios in wireless fading channels with feedback information \cite{tang}, interference and delay constrained cognitive radio relay channels \cite{musavian2010effective}, multiple-input multiple-output channels \cite{gursoy}, and multi-band cognitive radio channels \cite{elalemmoh}. Moreover, the authors in \cite{ozcan2014qos} studied the effective capacity of point-to-point channels and derived the optimal power allocation policies to maximize the system throughput by employing arbitrary input distributions under an average power constraint. More recently, we explored the effective capacity regions of multiple access channels with arbitrary input distributions and identified the optimal power allocation policies under average transmission power constraints \cite{hammouda2015effective}.

In this paper, different than the aforementioned studies and our recent study \cite{hammouda2015effective}, we focus on a broadcast channel scenario in which one transmitter employs arbitrarily distributed input signaling to convey data to two receivers under average power constraints and QoS requirements. We define the effective capacity region and provide an algorithm to obtain the optimal power allocation policies that maximize this region by enforcing the relation between the mutual information and the MMSE. Then, we express the optimal decoding regions in the space spanned by the channel fading power values. We finally justify our analytical results with numerical presentations.

\begin{figure}
\centering 
\includegraphics[width=0.84\textwidth]{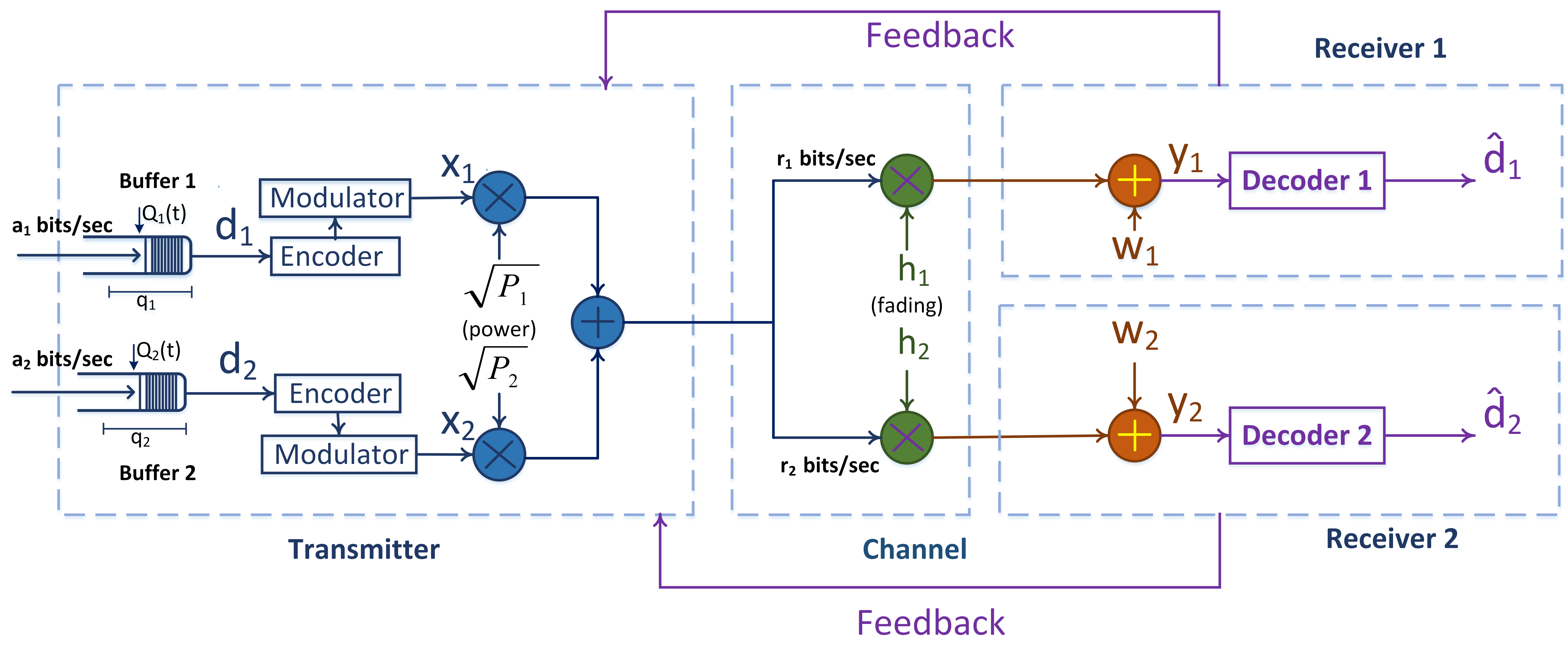}
\caption{Channel model: A two-user broadcast channel in which one transmitter communicates with two receivers. The transmitter performs superposition coding, while each receiver performs successive interference cancellation with a certain order. The decoding order depends on the channel conditions, i.e., the magnitude of the squares of the channel fading coefficients, $z_1 = |h_1|^2$ and $z_2 = |h_2|^2$.}\label{System_Model}
\end{figure}

\section{System Description}\label{sec:system_description}

\subsection{Channel Model}
As shown in Figure \ref{System_Model}, we consider a broadcast channel scenario in which one transmitter communicates with two receivers. We assume that the transmitter is equipped with two data buffers each of which stores data to be transmitted to the corresponding receiver. The transmitter, regarding the instantaneous channel conditions and employing a superposition coding strategy with a given order, sends data from both buffers in frames of $T$ seconds. During data transmission, the input-output relation between the transmitter and the $j^\text{th}$ receiver at time instant $t$ is given by
\begin{equation}\label{in_out_general}
y_{j}(t) =h_j(t)x_{j}(t)\sqrt{P_{j}(t)}+h_j(t)x_{m}(t)\sqrt{P_{m}(t)}+ w_j(t)\quad\text{for }t = 1,2,\cdots,
\end{equation}
where $j,m\in\{1,2\}$ and $j\neq m$. Above, $x_{j}(t)$ and $x_{m}(t)$ are the channel inputs at the transmitter and carries information to the $j^\text{th}$ and $m^\text{th}$ receivers, respectively, and $y_j(t)$ is the channel output at the $j^\text{th}$ receiver. Moreover, $w_j(t)$ represents the additive thermal noise at the $j^\text{th}$ receiver, which is a zero-mean, circularly symmetric, complex Gaussian random variable with a unit variance, i.e., ${E}\{|w_j|^2\}=1$. The noise samples $\{w_j(t)\}$ are assumed to be independent and identically distributed. Meanwhile, $h_j(t)$ represents the fading coefficient between the transmitter and the $j^\text{th}$ receiver, where ${E}\{|h_{j}|^{2}\}<\infty$. The magnitude square of the fading coefficient is denoted by $z_j(t)$, i.e., $z_j(t) = |h_j(t)|^2$. We consider a block-fading channel and assume that the fading coefficients stay constant for a frame duration of $T$ seconds and change independently from one frame to another. We further assume that $h_1$ and $h_2$ are independent of each other and perfectly known to the transmitter and both receivers. Thence, the transmitter can adapt the transmission power policy and the transmission rate for each receiver accordingly. In addition, the transmission power at the transmitter is constrained as follows:
\begin{equation}\label{avg_constraint_combined}
\mathbb{E}_{t}\{P_1(t)\}+\mathbb{E}_{t}\{P_2(t)\}\leq\overline{P},
\end{equation}
where $P_1(t)$ and $P_2(t)$ are the instantaneous power allocation policies for the $1^{\text{st}}$ and $2^{\text{nd}}$ receivers, respectively, i.e., $\mathbb{E}_{x_{1}}\{x_{1}(t)\}\leq P_1(t)$ and $\mathbb{E}_{x_{2}}\{x_{2}(t)\}\leq P_2(t)$, and $\overline{P}$ is finite. We finally note that the available transmission bandwidth is $B$ Hz. In the rest of the paper, we omit the time index $t$ unless otherwise needed for clarity.

\subsection{Achievable Rates}
In this section, we provide the instantaneous achievable rates between the transmitter and the receivers given the input signal distributions. We can express the instantaneous achievable rate between the transmitter and the $j^{\text{th}}$ receiver by invoking the mutual information between the channel inputs at the transmitter and the channel output at the $j^{\text{th}}$ receiver. Given that $h_{j}$ and $h_{m}$ are available at the transmitter and the $j^{\text{th}}$ receiver and that the $j^{\text{th}}$ receiver does not perform successive interference cancellation, the instantaneous achievable rate is given as \cite{gallager1968information}
\begin{equation*}
\mathcal{I}(x_{j};y_j)=\mathbb{E}\left\{\log_{2}\frac{f_{y_j|x_{j}}(y_j|x_{j})}{f_{y_j}(y_j)}\right\}\quad\text{for }j \in \{1,2\},
\end{equation*}
where $f_{y_j}(y_j)=\sum_{x_{j}}p_{x_{j}}(x_{j})f_{y_j|x_{j}}(y_j|x_{j})$ is the marginal probability density function (pdf) of the received signal $y_j$ and $f_{y_j|x_{j}}(y_j|x_{j})=\frac{1}{\pi} e^{-|y_j-h_jx_{j}\sqrt{P_{j}}|^2 }$. On the other hand, if the $j^{\text{th}}$ receiver performs successive interference cancellation, i.e., the $j^{\text{th}}$ receiver initially decodes $x_{m}$ and then decodes its own data, we have the achievable rate as follows:
\begin{equation*}
\mathcal{I}(x_{j};y_j|x_{m})=\mathbb{E}\left\{\log_{2}\frac{f_{u_{j}|x_{j}}(u_j|x_{j})}{ f_{u_j}(u_j)}\right\},
\end{equation*}
where $u_{j}=y_j-h_{j}x_{m}\sqrt{P_{m}}$. Above, $f_{u_j}(u_j)=\sum_{x_{j}}p_{x_{j}}(x_{j})f_{u_j|x_{j}}(u_j|x_{j})$ is the marginal pdf of $u_j$ and $f_{u_j|x_{j}}(u_j|x_{j})=\frac{1}{\pi} e^{-|u_j-h_jx_{j}\sqrt{P_{j}}|^2 }$.

We assume that each receiver, regarding the channel conditions and the encoding strategy at the transmitter, performs successive interference cancellation with a certain order if it is possible to do so. For instance, if the decoding order is ($j,m$) for $j,m \in \{1,2\}$ and $j \neq m$, the $j^{\text{th}}$ receiver decodes its own data by treating the signal carrying information to the $m^{\text{th}}$ receiver as interference. On the other hand, the $m^{\text{th}}$ receiver initially decodes the data sent to the $j^{\text{th}}$ receiver and subtracts the encoded signal from the channel output, and then decodes its own signal. Recall that both receivers perfectly know the instantaneous channel fading coefficients, $h_{1}$ and $h_{2}$, and the decoding order depends on the relation between the magnitude squares of channel fading coefficients $z_{1}$ and $z_{2}$. Therefore, we consider $\mathcal{Z}$ as the region in the $(z_1,z_2)$-space where the decoding order is (2,1) and $\mathcal{Z}^{c}$, the complement of $\mathcal{Z}$, as the region where the decoding order is (1,2). Noting that the transmitter can set the transmission rates to the instantaneous achievable rates, we can express the instantaneous transmission rate for the $1^{\text{st}}$ receiver as
\begin{align}
&r_1(z_1,z_2)=
\begin{cases}
\mathcal{I}(x_1;y_1|x_{2}),& \mathcal{Z},\\
\mathcal{I}(x_1;y_1), &\mathcal{Z}^{c},
\end{cases}\label{R_1}
\intertext{and the instantaneous transmission rate for the $2^{\text{nd}}$ receiver as}
&r_2(z_1,z_2)=
\begin{cases}
\mathcal{I}(x_2;y_2), &\mathcal{Z},\\
\mathcal{I}(x_2;y_2|x_{1}),&\mathcal{Z}^{c}.
\end{cases}\label{R_2}
\end{align}
The decoding regions can be determined in such a way to maximize the objective throughput.

\subsection{Effective Capacity}
Recall that the transmitter holds the data initially in the buffers. As a result, delay and buffer overflow concerns become of interest. Therefore, focusing on the data arrival processes at the transmitter, $a_1$ and $a_2$ in Fig. \ref{System_Model}, we invoke effective capacity as the performance metric. Effective capacity provides the maximum constant data arrival rate that a given service (channel) process can sustain to satisfy certain statistical QoS constraints \cite{wu_negi}. Let $Q$ be the stationary queue length at any data buffer. Then, we can define the decay rate of the tail distribution of the queue length $Q$ as 
\begin{equation*}
\theta=-\lim_{q\to\infty}\frac{\log_{e}\text{Pr}(Q\geq q)}{q}.
\end{equation*}
Hence, for a large threshold $q_{\max}$, we can approximate the buffer overflow probability as $\text{Pr}(Q\geq q_{\max})\approx e^{- \theta q_{\max}}$. Larger $\theta$ implies stricter QoS constraints, whereas smaller $\theta$ corresponds to looser constraints. For a discrete-time, stationary and ergodic stochastic service process $r(t)$, the effective capacity at the buffer is expressed as
\begin{equation*}
-\lim_{t \to \infty} \frac{1}{\theta t}\log_e\mathbb{E}\{e^{-\theta S(t)}\},
\end{equation*}
where $S(t) = \sum_{\tau = 1}^{t}r(\tau)$.

Since the transmitter in the aforementioned model has two different transmission buffers, we assume that each buffer has its own QoS requirements. Therefore, we denote the QoS exponent for each queue by $\theta_j$ for $j\in\{1,2\}$. Noting that the transmission bandwidth is $B$ Hz, the block duration is $T$ seconds, and the channel fading coefficients change independently from one transmission frame to another, we can express the effective capacity at each buffer in bits/sec/Hz as
\begin{equation}\label{effective_capacity_j}
a_{j}=-\frac{1}{\theta_{j}TB}\log_e\mathbb{E}\left\{e^{-\theta_{j}TBr_j(z_1,z_2)}\right\},
\end{equation}
where the expectation is taken over the space spanned by $z_1$ and $z_2$. Now, utilizing the definition given in \cite{qiao2013achievable}, we express the effective capacity region of the given broadcast transmission scenario as follows:
\begin{align}\label{effective_rate_region}
\mathcal{C}_E(\Theta)=&\bigcup_{r_1,r_2}\Big\{{C(\Theta)}\geq {\bf 0}:{C_j(\theta_{j})\leq a_{j}}\Big\},
\end{align}
where $\Theta = [\theta_1,\theta_2]$ is the vector of decay rates, ${C(\Theta)}=[C_{1}(\theta_{1}),C_{2}(\theta_2)]$ is the vector of the arrival rates at the transmitter buffers, and $\bf 0$ is the vector of zeroes.

\section{Performance Analysis}
In this section, we concentrate on maximizing the effective capacity region defined in (\ref{effective_rate_region}) under the QoS requirements for each transmitter buffer and the total average power constraint given in (\ref{avg_constraint_combined}). Notice that the effective capacity region is convex \cite{qiao2013achievable}. Hence, we can reduce our objective to maximizing the boundary surface of the region and express it as follows \cite{tse1998multiaccess}:
\begin{equation}\label{obtimization_objective}
\max_{\substack{\mathcal{Z},\mathcal{Z}^{c}\\\mathbb{E}\{P_{1}\}+\mathbb{E}\{P_{2}\}\leq\overline{P}}} \lambda_1 a_{1} + \lambda_2 a_{2},
\end{equation}
where $\lambda_1,\lambda_2\in[0,1]$ and $\lambda_1+\lambda_2=1$. In order to solve this optimization problem, we first obtain the power allocation policies in defined decoding regions, $\mathcal{Z}$ and $\mathcal{Z}^{c}$, and then, we provide the optimal decoding regions.

\subsection{Optimal Power Allocation}
Here, we derive the optimal power allocation policies that maximize the effective capacity region (\ref{obtimization_objective}) given $\mathcal{Z}$ and $\mathcal{Z}^c$. In the following analysis, we provide the proposition that gives the optimal power allocation policies:
\begin{proposition}\label{proposition_1}
The optimal power allocation policies, $P_1$ and $P_2$, that maximize the expression in (\ref{obtimization_objective}) are the solutions of the following equalities:
\begin{align}
&\frac{\lambda_1}{\psi_1}e^{-\theta_1 TBr_1(\textbf{z})} \frac{dr_1(\textbf{z})}{d P_1} + \frac{\lambda_2}{\psi_2} e^{-\theta_2 TBr_2(\textbf{z})} \frac{dr_2(\textbf{z})}{d P_1}  = \varepsilon, \label{optimal_alpha_1_Z}\\
&\frac{\lambda_2}{\psi_2} e^{-\theta_2 TBr_2(\textbf{z})} \frac{dr_2(\textbf{z})}{d P_2}  = \varepsilon, \label{optimal_alpha_2_Z}
\end{align}
for $\textbf{z} =[z_1,z_2] \in \mathcal{Z}$, and
\begin{align}
&\frac{\lambda_1}{\psi_1} e^{-\theta_1 TBr_1(\textbf{z})} \frac{dr_1(\textbf{z})}{d P_1} = \varepsilon, \label{optimal_alpha_1_Z_c}\\
&\frac{\lambda_1}{\psi_1} e^{-\theta_1 TBr_1(\textbf{z})} \frac{dr_1(\textbf{z})}{d P_2} + \frac{\lambda_2}{\psi_2} e^{-\theta_2 TBr_2(\textbf{z})} \frac{dr_2(\textbf{z})}{d P_2} = \varepsilon, \label{optimal_alpha_2_Z_c}
\end{align}
for $\textbf{z} \in \mathcal{Z}^c$. Above, $\psi_1 = \mathbb{E}_\textbf{z}\big \{ e^{-\theta_1 TBr_1(\textbf{z})} \big \}$ and $\psi_2 = \mathbb{E}_\textbf{z}\big\{ e^{-\theta_2 TBr_2(\textbf{z})} \big \}$, and $\varepsilon$ is the Lagrange multiplier of the average power constraint in (\ref{avg_constraint_combined}).
\end{proposition}
\begin{proof} Omitted due to the page limitation. \end{proof}

In Proposition \ref{proposition_1}, the derivatives of the transmission rates with respect to the corresponding power allocation policies are given as
\begin{align*}%\label{R_derivative}
\frac{dr_1(\textbf{z})}{d P_1}=\begin{cases}
\frac{d\mathcal{I}(x_1;y_1|x_{2})}{d P_1},& \mathcal{Z},\\
\frac{d\mathcal{I}(x_1;y_1)}{d P_1}, &\mathcal{Z}^{c},
\end{cases}\quad\text{and}\quad
\frac{dr_2(\textbf{z})}{d P_2}=\begin{cases}
\frac{d\mathcal{I}(x_2;y_2)}{d P_2}, &\mathcal{Z},\\
\frac{d\mathcal{I}(x_2;y_2|x_{1})}{d P_2},&\mathcal{Z}^{c},
\end{cases}
\end{align*}
and 
\begin{equation*}
\frac{dr_m(\textbf{z})}{d P_j} = \frac{d \mathcal{I}(x_j;y_j)}{d P_j} - \frac{d \mathcal{I}(x_j;y_j|x_{m})}{d P_j}\quad\text{for }m,j\in\{1,2\}\text{ and }m\neq j.
\end{equation*}

In the following theorem, we provide the derivatives of the mutual information with respect to the power allocation policies:
\begin{theorem}\label{theo:der_alpha}
Let $z_1$ and $z_2$ be given. The first derivative of the mutual information between $x_{j}$ and $y_j$ with respect to the power allocation policy, $P_j$, is given by
\begingroup
\allowdisplaybreaks
\begin{align}
&\frac{d \mathcal{I}(x_j;y_j)}{d P_j}=z_j \text{MMSE}(x_j;y_j)+z_j \sqrt{\frac{P_m}{P_j}} \text{Re}(\mathbb{E}\{x_j x_m^{*}-\hat{x}_j(y_{j})\hat{x}^{*}_m((y_{j}))\})
\label{d_I_tot_alpha}
\end{align}
for $j,m \in \{1,2\}$ and $j \neq m$. Above, $(\cdot)^{*}$ is the complex conjugate operation and $\text{Re}(\cdot)$ is the real part of a complex number. Meanwhile, the derivative of the mutual information between $x_{j}$ and $y_{j}$ with respect to $P_j$ given $x_{m}$ is
\begin{equation}
\frac{d\mathcal{I}(x_j;y_j|x_{m})}{d P_j}=z_j\text{MMSE}(x_j;y_j|x_{m}).\label{d_I_tot_alpha_2}
\end{equation}
\endgroup
MMSE and MMSE estimate are defined as
\begin{align*}
\text{MMSE}(u;v|s)=1-\frac{1}{\pi}\int\frac{\big|\sum_{u}u p(u) f_{v|u,s}(v|u,s)\big|^2}{f_{v|s}(v|s)} \mathrm{d}v
\end{align*}
and $\hat{u}(v)=\frac{\sum_{u}up(u)f_{v|u}(v|u)}{f_{v}(v)}$, respectively.
\end{theorem}
\begin{proof} Omitted due to the page limitation. \end{proof}

As clearly noticed in (\ref{optimal_alpha_1_Z})-(\ref{optimal_alpha_2_Z_c}), a closed-form solution for $P_{1}$ or $P_{2}$ cannot be obtained easily, which is mainly due to the tied relation between $P_{1}$ and $P_{2}$. For instance, $P_1$ is a function of $P_{2}$ as observed in (\ref{optimal_alpha_1_Z}) for $z \in \mathcal{Z}$, whereas $P_{2}$ is a function of $P_{1}$ as seen in (\ref{optimal_alpha_2_Z_c}) for $z \in \mathcal{Z}^c$. Therefore, we need to employ numerical techniques that consist of iterative solutions. Hence, in the following, we carry out an iterative algorithm that provides the optimal power allocation policies given decoding regions:

\begingroup\captionof{algorithm}{}\label{algorithm_1}
	\begin{algorithmic}[1] 
    \State Given $\lambda_1$, $\lambda_2$, $\mathcal{Z}$ and $\mathcal{Z}^c$;
    \State Initialize $\psi_1$, $\psi_2$;\label{int_psi}
    \While {True} 
    \State Initialize $\varepsilon$;
    \State Initialize $P_1$; \label{kappa}
    \While {True}
    \If {$z \in \mathcal{Z}$}
    \State For given $P_1$, compute the optimal $P_2$ by solving (\ref{optimal_alpha_2_Z}) \label{alg:alpha_2_Z};
    \State With obtained $P_2$, compute the optimal $P_1^{\star}$ by solving (\ref{optimal_alpha_1_Z}) \label{alg:alpha_1_Z};
    \Else 
    \State For given $P_1$, compute the optimal $P_2$ by solving (\ref{optimal_alpha_2_Z_c}) \label{alg:alpha_2_Z_c};
    \State For computed $P_2$, compute the optimal $P_1^{\star}$ by solving (\ref{optimal_alpha_1_Z_c}) \label{alg:alpha_1_Z_c};
    \EndIf
    \If {$|P_1 - P_1^{\star}| \leq \epsilon$ for small $\epsilon > 0$}
    \State break;
    \Else
    \State Set $P_1 = P_1^{\star}$;
    \EndIf
    \EndWhile
    \State Check if the average power constraint in (\ref{avg_constraint_combined}) is satisfied with quality;
    \State If not, update $\varepsilon$ and return to Step \ref{kappa}
    \State Compute $\psi_1^{\star} = \mathbb{E}_\textbf{z} \big \{ e^{-\theta_1 n r_1(\textbf{z})} \big \}$ and  $\psi_2^{\star} = \mathbb{E}_\textbf{z} \big \{ e^{-\theta_2 n r_2(\textbf{z})} \big \}$
    \If {$|\psi_1 - \psi_1^{\star}| \leq \epsilon$ and $|\psi_2 - \psi_2^{\star}| \leq \epsilon$}
    \State {\bf break};
    \Else
    \State Set $\psi_1 = \psi_1^{\star}$ and $\psi_2 = \psi_2^{\star}$;
    \EndIf
    \EndWhile
	\end{algorithmic}
\endgroup

Given $\lambda_j$ and $\psi_j$ for $j \in \{1,2\}$, it is shown in \cite{ozcan2014qos} that both (\ref{optimal_alpha_2_Z}) and (\ref{optimal_alpha_1_Z_c}) has at most one solution. We can further show that (\ref{optimal_alpha_1_Z}) has at most one solution for $P_{1}$ when $P_{2}$ is given, and that (\ref{optimal_alpha_2_Z_c}) has at most one solution for $P_{2}$ when $P_{1}$ is given. Consequently, we can guarantee that Steps \ref{alg:alpha_2_Z}, \ref{alg:alpha_1_Z}, \ref{alg:alpha_2_Z_c} and \ref{alg:alpha_1_Z_c} in Algorithm \ref{algorithm_1} will converge to a single unique solution. In addition, it is clear that (\ref{optimal_alpha_1_Z}) and (\ref{optimal_alpha_1_Z_c}) are monotonically decreasing functions of $P_1$, and that (\ref{optimal_alpha_2_Z}) and (\ref{optimal_alpha_2_Z_c}) are monotonically decreasing functions of $P_2$. Hence, in region $\mathcal{Z}$, we first obtain $P_{2}$ by solving (\ref{optimal_alpha_2_Z}) for given $P_{1}$, and then we find $P_{1}$ by solving (\ref{optimal_alpha_1_Z}) after inserting $P_{2}$ into (\ref{optimal_alpha_1_Z}). Similarly, in region $\mathcal{Z}^c$, we first obtain $P_{2}$ by solving (\ref{optimal_alpha_2_Z_c}) for given $P_{1}$, and then we find $P_{1}$ by solving (\ref{optimal_alpha_1_Z_c}) after inserting $P_{2}$ into (\ref{optimal_alpha_1_Z_c}). We can employ bisection search methods to obtain $P_{1}$ and $P_{2}$. In the above approach, when either $P_{1}$ or $P_{2}$ becomes negative, we set it to zero.

\subsection{Optimal Decoding Order}
Obtaining the optimal power allocation policies, we investigate the optimal decoding regions in this section. We initially notice that with no QoS constraints, i.e., $\theta_1 = \theta_2 = 0$, the effective capacity region is reduced to the ergodic capacity region. In this case, the symbol of the receiver with the strongest channel is always decoded last \cite{jindal2004duality}. Specifically, when $z_j \geq z_m$ the symbol of the $j^\text{th}$ receiver is decoded last. This result is based on the assumption of Gaussian input signaling. To the best of our knowledge, no such a result is obtained for broadcast channels when QoS constraints are applied, i.e, $\theta_1 >0$ and $\theta_2 > 0$, and/or when arbitrary input signaling is employed. In the following, we consider a special case of $\theta_1 = \theta_2$ for $\theta > 0$ and provide the optimal decoding order regions when arbitrary input distributions are employed by the transmitter:

\begin{theorem}\label{theo:optimal_order}
Let $z_1$, $z_2$ and $\overline{P}$ be given. Define $z_1^\star$ for any given $z_2\geq 0$ such that the decoding order is (2,1) when $z_2>z_1^{\star}$. Otherwise, it is (1,2). With arbitrary input distributions and power allocation policies at the transmitter, the optimal $z_1^\star$ for any given $z_2$ value is the solution of the following equality:
\begin{equation}\label{optimal_oeder}
\mathcal{I}(x_{1},x_{2};y_{1},y_{2}|z_1^\star,z_2) = \mathcal{I}(x_1;y_1|x_{2},z_1^\star) + \mathcal{I}(x_2;y_2|x_{1},z_{2}).
\end{equation}
\end{theorem}
\begin{proof} Omitted due to the page limitation. \end{proof}

\begin{figure}
\centering 
\includegraphics[width=0.6\textwidth]{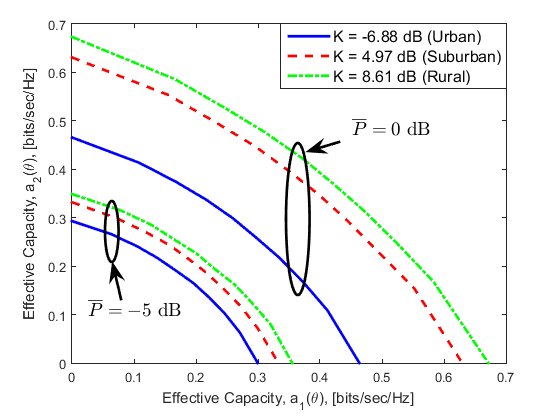}
\caption{Effective capacity region boundary when BPSK input signaling is employed for different values of $\overline{P}$ and $K$. The areas under the curves provide the effective capacity regions.}\label{BC_K}
\end{figure}

\section{Numerical Results}
In this section, we present the numerical results. Throughout the paper, we set the available channel bandwidth to $B = 100$ Hz and the transmission frame duration to $T = 1$ second. We further assume that $h_1$ and $h_2$ are independent of each other and set $\mathbb{E}\{|h_1|^2\} = \mathbb{E}\{|h_2|^2\} = 1$. In addition, we assume that signals transmitted to both receivers are independent of each other, i.e., $\mathbb{E}\{x_j x_m^{*}\} = 0$. Unless indicated otherwise, we set the QoS exponents $\theta_1 = \theta_2 = 0.01$. We finally assume that both receivers have the same noise statistics, i.e., $\mathbb{E}\{|w|^2\}= \mathbb{E}\{|w_1|^2\} = \mathbb{E}\{|w_2|^2\} = 1$, and we define the received signal-to-noise ratio at each receiver with $\frac{\overline{P}}{{E}\{|w|^2\}}=\overline{P}$. 

\begin{figure}
\centering 
\includegraphics[width=0.6\textwidth]{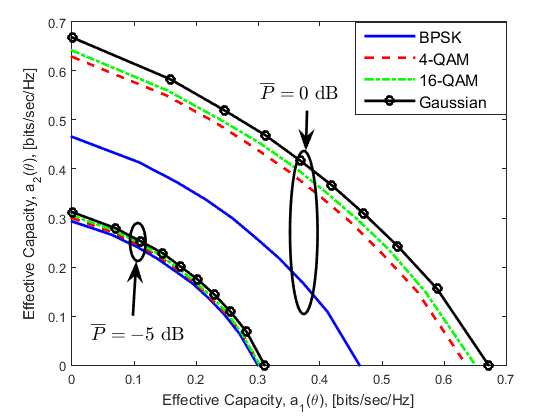}
\caption{Effective capacity region boundary with different modulation techniques and signal-to-noise ratio values when $K = -6.88$ dB.}\label{BC_Mod}
\end{figure}

In Fig. \ref{BC_K}, we initially consider binary phase shift keying (BPSK) employed at the transmitter for both receivers and investigate the effect of channel statistics on the effective capacity region in Rician fading channels with a line-of-sight parameter $K$, which is the ratio of the power in the line-of-sight component to the total power in the non-line-of-sight components. The empirical values of $K$ are determined to be -6.88 dB, 8.61 dB and 4.97 dB for urban, rural and suburban environments at 781 MHz, respectively \cite{jeong2012mimo}. Considering these $K$ values, we obtain results for different signal-to-noise ratio values, i.e., $\overline{P}=0$ and $-5$ dB. We can clearly see that the effective capacity region broadens as $K$ increases because the line-of-sight component becomes more dominant with increasing $K$. We also observe that the effect of $K$ is more apparent when the signal-to-noise ratio is greater. 

\begin{figure}
\centering 
\includegraphics[width=0.6\textwidth]{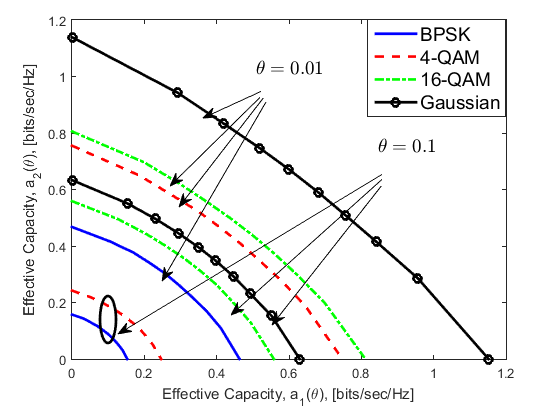}
\caption{Effective capacity region boundary with different modulation techniques and decay rate parameters, $\theta=\theta_{1}=\theta_{2}$ when $K = -6.88$ dB and $\overline{P} = 5$ dB.}\label{BC_theta}
\end{figure}

Subsequently, setting $K=-6.88$ dB, we investigate the effect of different signal modulation techniques with different signal-to-noise ratio values in Fig. \ref{BC_Mod}. We consider BPSK, quadrature amplitude modulation (QAM) and Gaussian input signaling. We can easily notice the superiority of the Gaussian input signaling over the others, while BPSK has the lowest performance. However, the performance gap is reduced with decreasing $\overline{P}$. Lastly, we explore the effect of the QoS exponent $\theta$ on the effective capacity performance in Fig. \ref{BC_theta}. We set $\overline{P} = 5$ dB and $K = -6.878$ dB and plot results for different modulation techniques. Increasing $\theta$ results in a smaller effective capacity region as the system is subject to stricter QoS constraints. 

\section{Conclusion}
In this paper, we have examined optimal power allocation policies that maximize the effective capacity region of a two-user broadcast transmission scenario with arbitrarily distributed input signals. We have invoked the relation between MMSE and the first derivative of mutual information with respect to transmission power. We have proposed an iterative algorithm that converges to optimal power allocation policies given decoding regions under an average power constraint. Obtaining power allocation policies, we have further characterized decoding regions for successive interference cancellation by the receivers. Through numerical solutions, we have substantiated our results. In general, there is an apparent superiority of Gaussian input signaling over the other modulation techniques, whereas the gap between Gaussian input signaling and the others decreases with decreasing signal-to-noise ratio. Therefore, it is reasonable to employ simple modulation techniques in low signal-to-noise ratio regimes.

\bibliographystyle{splncs03}
\bibliography{IEEEfull,references_final}

\begin{thebibliography}{10}
\providecommand{\url}[1]{\texttt{#1}}
\providecommand{\urlprefix}{URL }

\bibitem{ak_gur_4}
Akin, S., Gursoy, M.: On the throughput and energy efficiency of cognitive
  {MIMO} transmissions. {IEEE} Transactions on Vehicular Technology  62(7),
  3245--3260 (September 2013)

\bibitem{bergmans1973random}
Bergmans, P.P.: Random coding theorem for broadcast channels with degraded
  components. {IEEE} Transactions on Information Theory  19(2),  197--207
  (1973)

\bibitem{cover1972broadcast}
Cover, T.M.: Broadcast channels. {IEEE} Transactions on Information Theory
  18(1),  2--14 (1972)

\bibitem{elalemmoh}
Elalem, M., Zhao, L.: Effective capacity and interference analysis in multiband
  dynamic spectrum sensing. Communications and Network  5(2),  111--118 (May
  2013)

\bibitem{gallager1968information}
Gallager, R.G.: Information theory and reliable communication. Springer (1968)

\bibitem{gallager1974capacity}
Gallager, R.G.: Capacity and coding for degraded broadcast channels. Problemy
  Peredachi Informatsii  10(3),  3--14 (1974)

\bibitem{gunduz2010gaussian}
G{\"u}nd{\"u}z, D., Payar{\'o}, M.: {Gaussian} two-way relay channel with
  arbitrary inputs. In: International Symposium on Personal Indoor and Mobile
  Radio Communications (PIMRC). pp. 678--683. IEEE (2010)

\bibitem{guo2005mutual}
Guo, D., Shamai, S., Verd{\'u}, S.: Mutual information and minimum mean-square
  error in {Gaussian} channels. {IEEE} Transactions on Information Theory
  51(4),  1261--1282 (April 2005)

\bibitem{gupta2006power}
Gupta, G., Toumpis, S., et~al.: Power allocation over parallel {Gaussian}
  multiple access and broadcast channels. {IEEE} Transactions on Information
  Theory  52(7),  3274--3282 (2006)

\bibitem{gursoy}
Gursoy, M.: {MIMO} wireless communications under statistical queueing
  constraints. {IEEE} Transactions on Information Theory  57(9),  5897--5917
  (September 2011)

\bibitem{hammouda2015effective}
Hammouda, M., Akin, S., Peissig, J.: Effective capacity in multiple access
  channels with arbitrary inputs. In: {IEEE} International Conference on
  Wireless and Mobile Computing, Networking and Communications (WiMob). pp.
  406--413 (2015)

\bibitem{jeong2012mimo}
Jeong, W.H., Kim, J.S., Jung, M.w., Kim, K.S.: {MIMO} channel measurement and
  analysis for {4G} mobile communication. In: Convergence and Hybrid
  Information Technology, pp. 676--682. Springer (2012)

\bibitem{jindal2004duality}
Jindal, N., Vishwanath, S., Goldsmith, A.: On the duality of {Gaussian}
  multiple-access and broadcast channels. {IEEE} Transactions on Information
  Theory  50(5),  768--783 (2004)

\bibitem{li2001capacity}
Li, L., Goldsmith, A.J.: Capacity and optimal resource allocation for fading
  broadcast channels. {I}. {E}rgodic capacity. {IEEE} Transactions on
  Information Theory  47(3),  1083--1102 (2001)

\bibitem{lozano2006optimum}
Lozano, A., Tulino, A.M., Verd{\'u}, S.: Optimum power allocation for parallel
  {Gaussian} channels with arbitrary input distributions. {IEEE} Transactions
  on Information Theory  52(7),  3033--3051 (2006)

\bibitem{musavian2010effective}
Musavian, L., A{\"\i}ssa, S., Lambotharan, S.: Effective capacity for
  interference and delay constrained cognitive radio relay channels. {IEEE}
  Transactions on Wireless Communications  9(5),  1698--1707 (May 2010)

\bibitem{nguyen2010outage}
Nguyen, K.D., Guillen~i Fabregas, A., Rasmussen, L.K.: Outage exponents of
  block-fading channels with power allocation. {IEEE} Transactions on
  Information Theory  56(5),  2373--2381 (2010)

\bibitem{ozcan2014qos}
Ozcan, G., Gursoy, M.C.: {QoS}-driven power control for fading channels with
  arbitrary input distributions. In: {IEEE} International Symposium on
  Information Theory (ISIT). pp. 1381--1385. IEEE (2014)

\bibitem{qiao2013achievable}
Qiao, D., Gursoy, M.C., Velipasalar, S.: Achievable throughput regions of
  fading broadcast and interference channels under {QoS} constraints. {IEEE}
  Transactions on Communications  61(9),  3730--3740 (2013)

\bibitem{tang}
Tang, J., Zhang, X.: Quality-of-service driven power and rate adaptation over
  wireless links. {IEEE} Transactions on Wireless Communications  6(8),
  3058--3068 (August 2007)

\bibitem{tao2012overview}
Tao, X., Xu, X., Cui, Q.: An overview of cooperative communications. {IEEE}
  Communications Magazine  50(6),  65--71 (2012)

\bibitem{tse1997optimal}
Tse, D.N.: Optimal power allocation over parallel {Gaussian} broadcast
  channels. In: {IEEE} International Symposium on Information Theory (ISIT).
  pp. 27--27. Citeseer (1997)

\bibitem{tse1998multiaccess}
Tse, D.N.C., Hanly, S.V.: Multiaccess fading channels. {I}. {P}olymatroid
  structure, optimal resource allocation and throughput capacities. {IEEE}
  Transactions on Information Theory  44(7),  2796--2815 (1998)

\bibitem{wu_negi}
Wu, D., Negi, R.: Effective capacity: A wireless link model for support of
  quality of service. {IEEE} Transactions on Wireless Communications  2(4),
  630--643 (July 2003)

\end{thebibliography}
\end{document}